\documentclass{article}

\usepackage{PRIMEarxiv}

\usepackage[utf8]{inputenc} 
\usepackage[T1]{fontenc}    
\usepackage{hyperref}       
\usepackage{url}            
\usepackage{booktabs}       
\usepackage{amsfonts}       
\usepackage{nicefrac}       
\usepackage{microtype}      
\usepackage{lipsum}
\usepackage{fancyhdr}       
\usepackage{graphicx}       
\graphicspath{{media/}}     

\usepackage{float}
\usepackage{booktabs}
\usepackage{listings}
\usepackage{subcaption}
\usepackage{mathtools}
\usepackage{relsize}
\usepackage[export]{adjustbox}

\usepackage{xcolor}

\colorlet{punct}{red!60!black}
\definecolor{background}{HTML}{ffffff}
\definecolor{delim}{RGB}{20,105,176}
\colorlet{numb}{magenta!60!black}

\lstdefinelanguage{json}{
    basicstyle=\footnotesize\ttfamily,
    columns=fullflexible,
    numbers=left,
    numberstyle=\scriptsize,
    numbersep=4pt,
    showstringspaces=false,
    breaklines=true,
    frame=lines,
    backgroundcolor=\color{background},
    literate=
     *{0}{{{\color{numb}0}}}{1}
      {1}{{{\color{numb}1}}}{1}
      {2}{{{\color{numb}2}}}{1}
      {3}{{{\color{numb}3}}}{1}
      {4}{{{\color{numb}4}}}{1}
      {5}{{{\color{numb}5}}}{1}
      {6}{{{\color{numb}6}}}{1}
      {7}{{{\color{numb}7}}}{1}
      {8}{{{\color{numb}8}}}{1}
      {9}{{{\color{numb}9}}}{1}
      {:}{{{\color{punct}{:}}}}{1}
      {,}{{{\color{punct}{,}}}}{1}
      {\{}{{{\color{delim}{\{}}}}{1}
      {\}}{{{\color{delim}{\}}}}}{1}
      {[}{{{\color{delim}{[}}}}{1}
      {]}{{{\color{delim}{]}}}}{1},
}

\title{GEOS: Scalability Analysis of a Global Blockchain for Immunization Records
}

\author{
  Jorge Medina, Roberto Rojas-Cessa \\
  New Jersey Institute of Technology \\
  Newark, NJ, USA\\
  \texttt{\{jorge.medina, rojas\}@njit.edu} \\
   \And
  Ziqian Dong \\
  New York Institute of Technology \\
  New York, NY, 10023, USA\\
  \texttt{ziqian.dong@nyit.edu} \\
   \AND
   Vatcharapan Umpaichitra\\
   SUNY Downstate Health Sciences University \\
   Brooklyn, NY, 11203 \\
   \texttt{vatcharapan.umpaichitra@downstate.edu} \\
}

\begin{document}
\maketitle

\begin{abstract}
While vaccinations continue to be rolled out to curb the ongoing COVID-19 pandemic, their verification is becoming a requirement for the re-incorporation of individuals into many social activities or travel. Blockchain technology has been widely proposed to manage vaccination records and their verification in many politically-bound regions. However, the high contagiousness of COVID-19 calls for a global vaccination campaign. Therefore, a blockchain for vaccination management must scale up to support such a campaign and be adaptable to the requirements of different countries. While there have been many proposals of blockchain frameworks that balance the access and immutability of vaccination records, their scalability, a critical feature, has not yet been addressed. 

In this paper, we propose a scalable and cooperative Global Immunization Information Blockchain-based System (GEOS) that leverages the global interoperability of immunization information systems. We model GEOS and describe its requirements, features, and operation. We analyze the communications and the delays incurred by the national and international consensus processes and blockchain interoperability in GEOS. Such communications are pivotal in enabling global-scale interoperability and access to electronic vaccination records for verification. We show that GEOS ably keeps up with the global vaccination rates of COVID-19 as an example of its scalability.
\end{abstract}


\section{Introduction}
A global vaccination campaign as an attempt to curb the COVID-19 pandemic is underway. COVID-19, caused by the severe acute respiratory syndrome coronavirus 2 (SARS-CoV-2), has threatened global health and inflicted economic hardship on millions of people in almost every corner of the globe. SARS-CoV-2 has proven to be remarkably infectious and deadly, and as time passes, it mutates into new strains and that further exacerbates the containment of COVID-19 ~{\cite{hu2021characteristics,bar2020science,abdool2021new,gomez2021emerging,wang2022sars}}.

At the time of writing this paper, the world has authorized multiple vaccines for their administration to the world population. These vaccines aim to prevent COVID-19, severe illness, and death {\cite{polack2020safety, baden2021efficacy,sadoff2021safety,voysey2021single}}. The vaccination rollout is being sped up to reduce opportunities for the virus to mutate into more infectious and morbid strains than the original one. For the sake of a fast vaccination campaign, it may be attractive to simplify vaccination management and record keeping. But loose management of the world’s vaccination against COVID-19 may undermine vaccination effectiveness and the enforcement of preventive measures that reduce cases and deaths, minimize opportunities for viral mutation and fully restore normal life {\cite{medina2021reducing,jiang2021energy}}.

Globalization not only empowers the world to become more interconnected but also magnifies the population's exposure to infectious diseases and renders local-scale campaigns insufficient to prevent recurrences of such a disease in the long term {\cite{bunnell2019global}}. COVID-19 is a clear example. Therefore, a coordinated global-scale vaccination campaign is being sought to bound the spread of COVID-19 {\cite{bassi2021allocating,andreadakis2020covid,yamey2020ensuring}}. By providing global access to information, electronic records can play a crucial role in the vaccination process and management of preventive measures. However, they must also ensure security, privacy, and standardization for global adoption.

Blockchain, an immutable ledger, has been pointed at as the technology of choice to realize accessible and secure electronic records {\cite{crosby2016blockchain,ramamurthy2020blockchain}}. The immutability property of blockchain is achieved using a distributed consensus algorithm where multiple and independent validator nodes in a network, or validators for short, agree on the addition of new data records. Validators communicate through a peer-to-peer (P2P) network and run the consensus algorithm to keep their blockchain copies synchronized.

Many blockchain solutions have been recently proposed to record, manage, and enable ubiquitous access to electronic vaccination records {\cite{nortey2019privacy,antal2021blockchain,yong2019intelligent,sharma2020blockchain}}. Such solutions aim to securely support the identification of vaccinated individuals and make them eligible for participating in social activities. 
Those solutions also aim to empower the implementation of preventive measures and facilitate international travel during a pandemic by using a reliable vaccination verification process.

While proposals for blockchain-based vaccination record management are abound, in the form of electronic records or immunization passports, it is unknown whether blockchain is scalable enough for recording the vaccination of the world’s population. A global blockchain of vaccination records must achieve a transaction rate large enough to sustain not only the world population but also the daily vaccination rates of the most populous countries. Such rates can be achieved by achieving short block confirmation times {\cite{goad2013vaccinations}}.
Equally important, such a global blockchain system must also enable interoperation of information among countries, where each country’s immunization information system may operate autonomously {\cite{world2007strengthening}}.

In this paper, we propose a scalable Global Immunization Information Blockchain-based System (GEOS) that leverages the global interoperability of immunization information records. GEOS enables the interoperability of national immunization information systems and, more critically, the scalability of a global blockchain-based immunization information system.
We model GEOS's structured P2P communication network to evaluate its performance in terms of block confirmation time for recording vaccination transactions. We show that GEOS's confirmation time can sustain global-scale vaccination rates to cover the world population and an administration of two doses per person per year, as currently recommended for many COVID-19 vaccines.
Our contributions in this paper are the proposal of a global-scale blockchain for domestic and international vaccination verification, a delay model of the communications performed in a global-scale blockchain, and extensive evaluations of our model to demonstrate that GEOS scales up with the maximum possible global demand of vaccinations.

This paper is organized as follows: Section~\ref{sec:GEOS} introduces our proposed global blockchain for vaccination and verification management GEOS. Section~\ref{sec:model} presents a mathematical model of the confirmation time of GEOS. Section~\ref{sec:performance} evaluates the scalability of GEOS for its use as a global electronic immunization record system. Section~\ref{sec:discussion} discusses existing challenges for a successful adoption of GEOS. Section~\ref{sec:literature-review} reviews the existing work in blockchain for immunization management. Section~\ref{sec:conclusions} presents our conclusions.

\section{Global Immunization Information Blockchain-based System}
\label{sec:GEOS}

\begin{figure*}[htp!]
    \centering
    \includegraphics[scale=0.17]{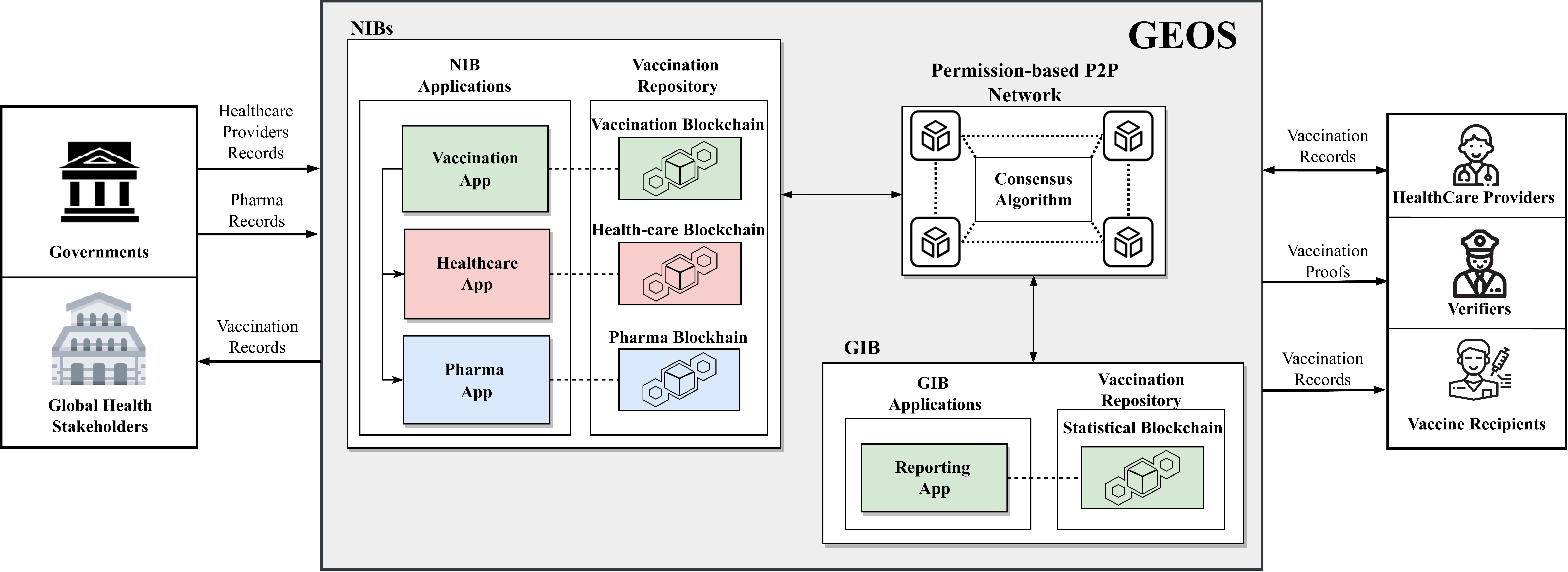}
    \caption{Components of the proposed GEOS. It comprises NIBs and a GIB. Each of them is accessible through applications that use smart contracts, and each NIB and GIB use a consensus algorithm to commit data to a distributed vaccination repository with blockchain ledgers over a permission-based P2P network.}
    \label{fig:GEOS-Components}
\end{figure*}

GEOS is a blockchain-based vaccination record system that stores vaccination records of individuals and global vaccination statistics in secure and standardized blockchain ledgers. It enables global access and sharing of immunization information by using a two-layered system, a lower layer for national vaccination records of a country, and an upper layer for international interoperability and global vaccination statistics. {Fig.~\ref{fig:GEOS-Components}} shows the components of GEOS and user accessibility. The lower layer of GEOS comprises National Immunization Blockchains (NIBs) and the upper layer, a Global Immunization Blockchain (GIB). GIB and NIBs are similar consortium blockchains {\cite{yao2021survey}} that serve as distributed vaccination repositories. GEOS interconnects NIBs and GIB through a permission-based P2P network of distributed validators.

NIBs keep the vaccination records of the population of a country and keep them accessible for the implementation of local preventive measures against COVID-19 or other infectious agents. NIBs' vaccination records support licensed healthcare providers' administration of legitimate vaccines and enable worldwide vaccination verification. GIB enforces the compliance of participating NIBs to record vaccination records and report vaccination statistics according to GEOS policies. It also serves as the global repository of vaccination statistics of NIBs. Vaccine recipients, healthcare providers, or vaccination verification officers may be NIB users. International institutions like WHO and UNICEF may be GIB users. NIBs and GIB record vaccination data using transactions. Validators in NIBs and GIB execute applications that use smart contracts and a consensus algorithm to verify and record transactions.

\subsection{Vaccination Repositories}

The vaccination repository of NIBs combines the records of two blockchains: 1) healthcare and 2) pharma blockchains, and the vaccination records as a vaccination blockchain. The healthcare and pharma blockchains store the data of authorized healthcare providers and the data of manufactured vaccines, respectively. The stored information in a pharma blockchain includes the manufacturer Id, serial number, manufacturing and expiration date of vaccines, and others. The healthcare blockchain keeps the name, Id, and authorization records of the healthcare providers. Validators in NIBs use a vaccination application to verify, add and query vaccination records. The vaccination repository of GIB comprises a blockchain that keeps and consolidates the vaccination statistics reported by participating NIBs. Validators in GIB use a reporting application to query, verify, and add the reported vaccination statistics.

\subsubsection{Vaccination Transactions: iRecords and iReports}
In NIBs, an authorized healthcare provider issues a transaction, or iRecord, to record a vaccination. An iRecord contains information such as the name of the vaccine recipient, the healthcare provider who administered the vaccine, vaccine name and serial number, and vaccine manufacturer. Validators keep iRecords stored in their vaccination blockchains and provide access to such data to authorized users. {Listing~\ref{listing::iRecord}} shows an example of the contents of an iRecord.

\begin{lstlisting}[linewidth=\columnwidth,language=json,numbers=none,caption={Example of contents in an iRecord.},label=listing::iRecord]
{
    "txId": "<Hash of transaction>",
    "timestamp": "<Time of message creation>",
    "recipient": {
        "Id": "<Vaccine recipient public key>",
        "cc": "<Country code>"
    },    
    "vaccine" :{
        "vaccode":"<Vaccine code>",
        "route": "<Route of administration>",
        "timestamp": "<Time of administration>"
        "dose": "<Dose's number>",
    },
    "manufacturer" :{
        "name" : "<Vaccine manufacturer's Id>",
        "cc" : "<Manufacturer's country code>",
        "vacname": "<Vaccine's name>",
        "serialnum" : "<Vaccine's serial number>",
        "batchnum" : "<Vaccine's batch number>"
    },    
    "healthcare" : { 
        "name": "<Healthcare provider's name>",
        "location": "<Healthcare provider's address>",
        "cc": "<Country code>"
    },
    "signerID" : "<The public key of healthcare provider>",
    "sig": "<Signature of healthcare provider>"
}
\end{lstlisting}

Each NIB sends a transaction of the vaccination statistics, or vaccination report (iReport), to GIB. The reported statistics may include the total number of administered vaccines, vaccine names, and eligible and anonymous information of vaccinated individuals. Validators in NIBs issue an iReport to GIB after committing a new block. GIB validators authenticate the NIB validators, verify the iReports, and commit them to their blockchain. {Listing~\ref{listing::iReports}} shows an example of the contents of an iReport.

\begin{lstlisting}[linewidth=\columnwidth,language=json,numbers=none,caption={Example of contents in an iReport.},label=listing::iReports]
{
    "txId": "<Hash of transaction>",
    "timestamp": "<Time of message creation>",
    "cc" : "<NIB's country code>",
    "report": ["<List of administered vaccinations split by vaccine name>"],
    "Q": {
        "QId": ["<List of signing NIB validators' membership Ids>"],
        "aggSig" : "< Quorum aggregated signature>"
    },
    "signerId" : "<The public key of the sender NIB validator>",
    "sig": "<Signature of sender NIB validator>"
}
\end{lstlisting}

\subsection{P2P Network}\label{p2p}
\begin{figure}[htp!]
    \centering
    \includegraphics[width=0.7\columnwidth]{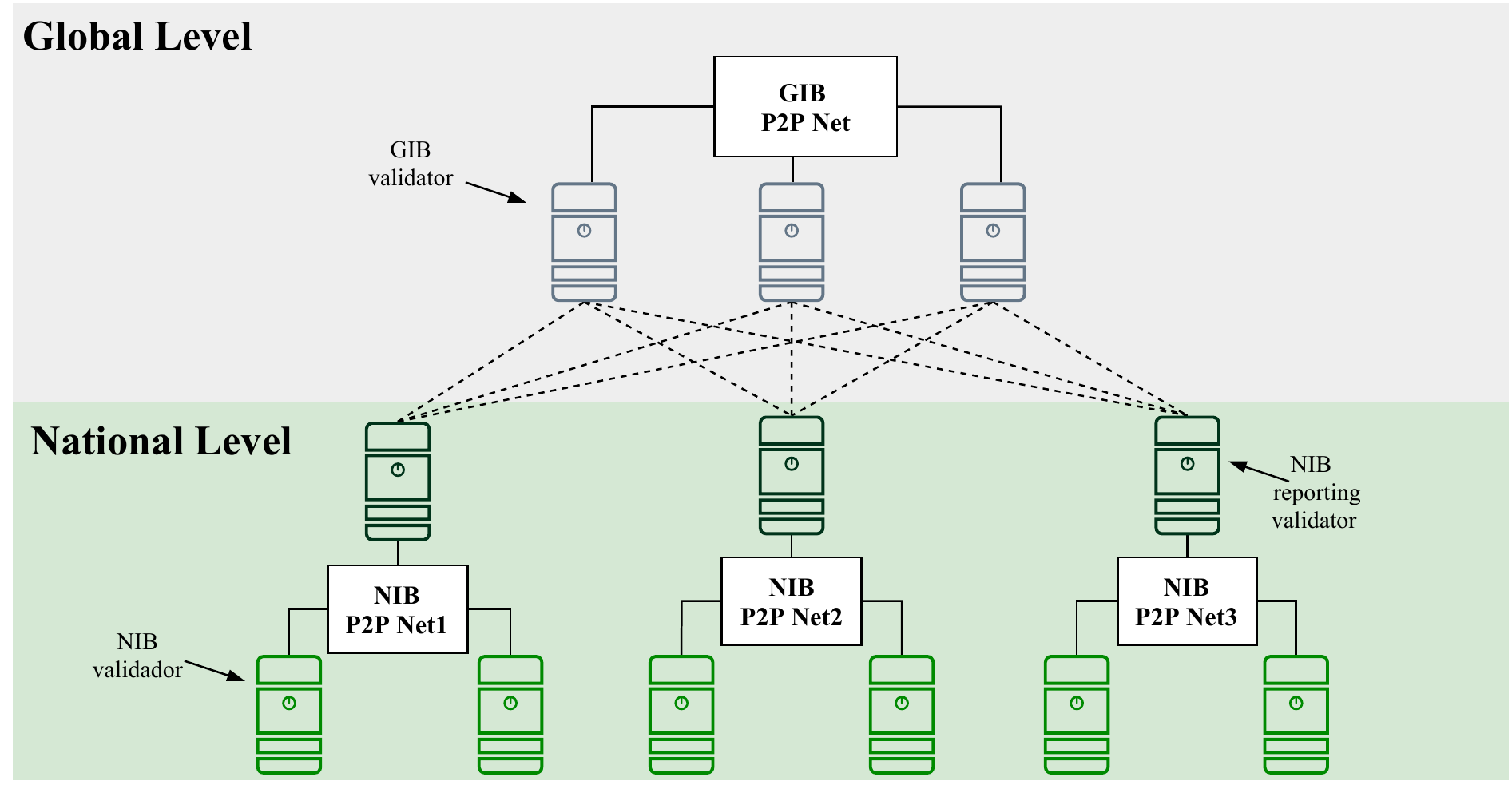}
    \caption{The architecture of the permission-based P2P network of validators used in GEOS. Validators in an NIB are distributed across the associated country, and validators in GIB are distributed across different countries. NIBs designate one or more validators to report vaccination statistics to GIB.}
    \label{fig:p2p}
\end{figure}

The architecture of the permission-based P2P networks in GEOS is structured in two levels: a global level for GIB and a national level for NIBs, with one P2P network per NIB, as {Fig.~\ref{fig:p2p}} shows. The global level has GIB validators distributed across different countries. The national level consists of one P2P network of validators per NIB, distributed across a country. One or more validators are designated to issue iReports for GIB in such a network.

NIBs and GIB P2P networks are structured overlay networks that communicate through the Internet using the Kademlia protocol ~{\cite{maymounkov2002kademlia}}. These networks also use the Kademlia protocol to implement the broadcast algorithm proposed in {\cite{czirkos2013solution, rohrer2019kadcast}} for broadcasting transactions, blocks, and messages associated with the consensus process. This broadcast algorithm reduces message complexity by enabling validators to share broadcast responsibility. 

\subsection{Consensus Process}
NIBs and GIB use the Hotstuff consensus algorithm to agree on adding a new block and consortium membership management~{\cite{yin2019hotstuff}}. The complexity of the consensus algorithm is a linear function of the number of validators. GEOS uses the Boneh-Lynn-Shacham (BLS) cryptographic signature scheme to aggregate the signatures from multiple validator votes {\cite{boneh2018compact}}. Such scheme reduces the number of messages exchanged during the consensus process. The scheme is based on having a leader validator, or leader for short, managing communications for verification with the other validators.

\begin{figure}[htp!]
    \centering
    \includegraphics[width=0.6\columnwidth]{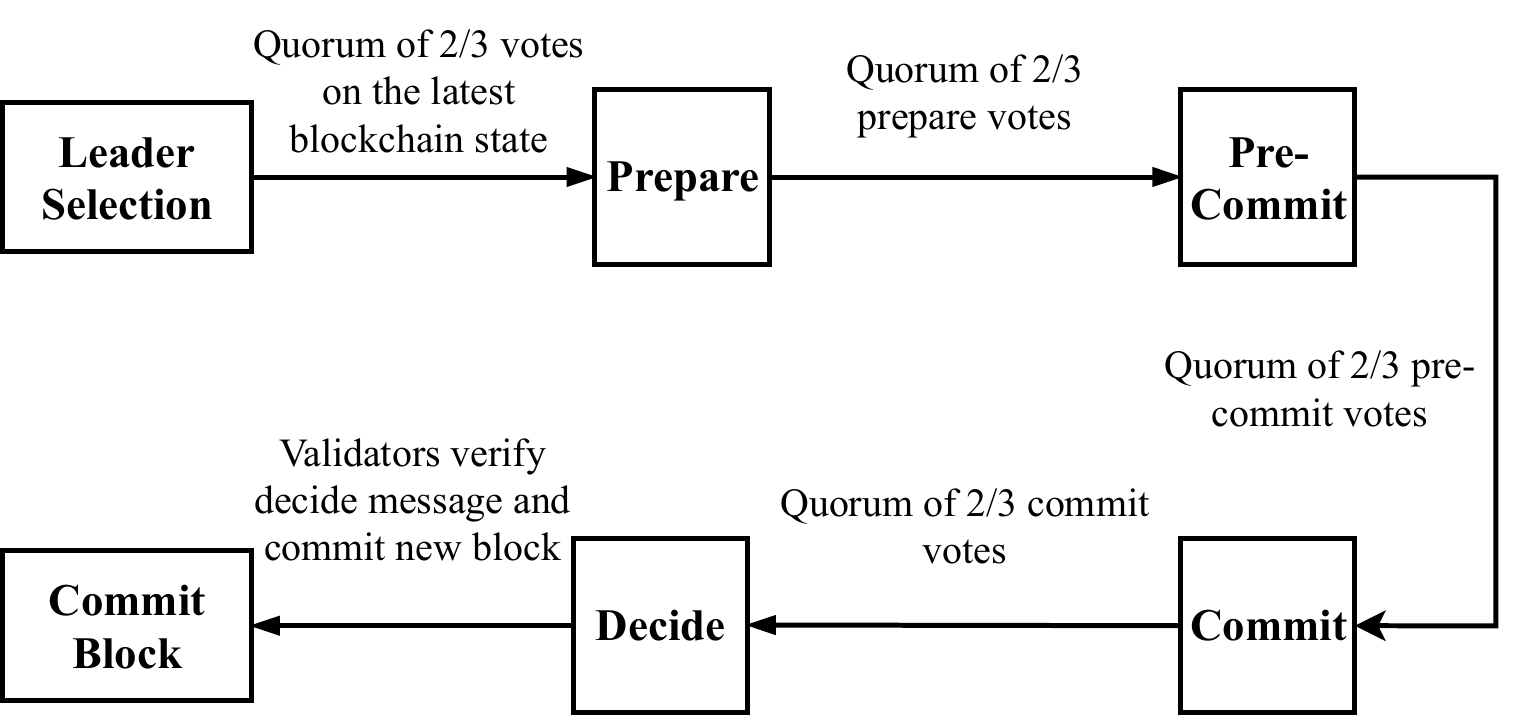}
    \caption{Phases of the consensus algorithm that NIBs and GIB use to agree on new vaccination data.}
    \label{fig::consensus-phases}
\end{figure}

The consensus algorithm operates in four phases: Prepare, Pre-commit, Commit, and Decide, as depicted in {Fig.~\ref{fig::consensus-phases}}. In every consensus run, the selection of a leader follows a round-robin schedule among all validators. In every phase, the leader broadcasts a quorum certificate, $Q$, to the validators. This certificate contains a list of the received votes in each event for each validator to verify. The validators send a vote as a response after verification. The consensus process moves from one phase to the next after the leader receives a quorum size, $q_s$, of 2/3 of validator votes.

\begin{figure}[htp!]
    \centering
    \includegraphics[width=0.60\columnwidth]{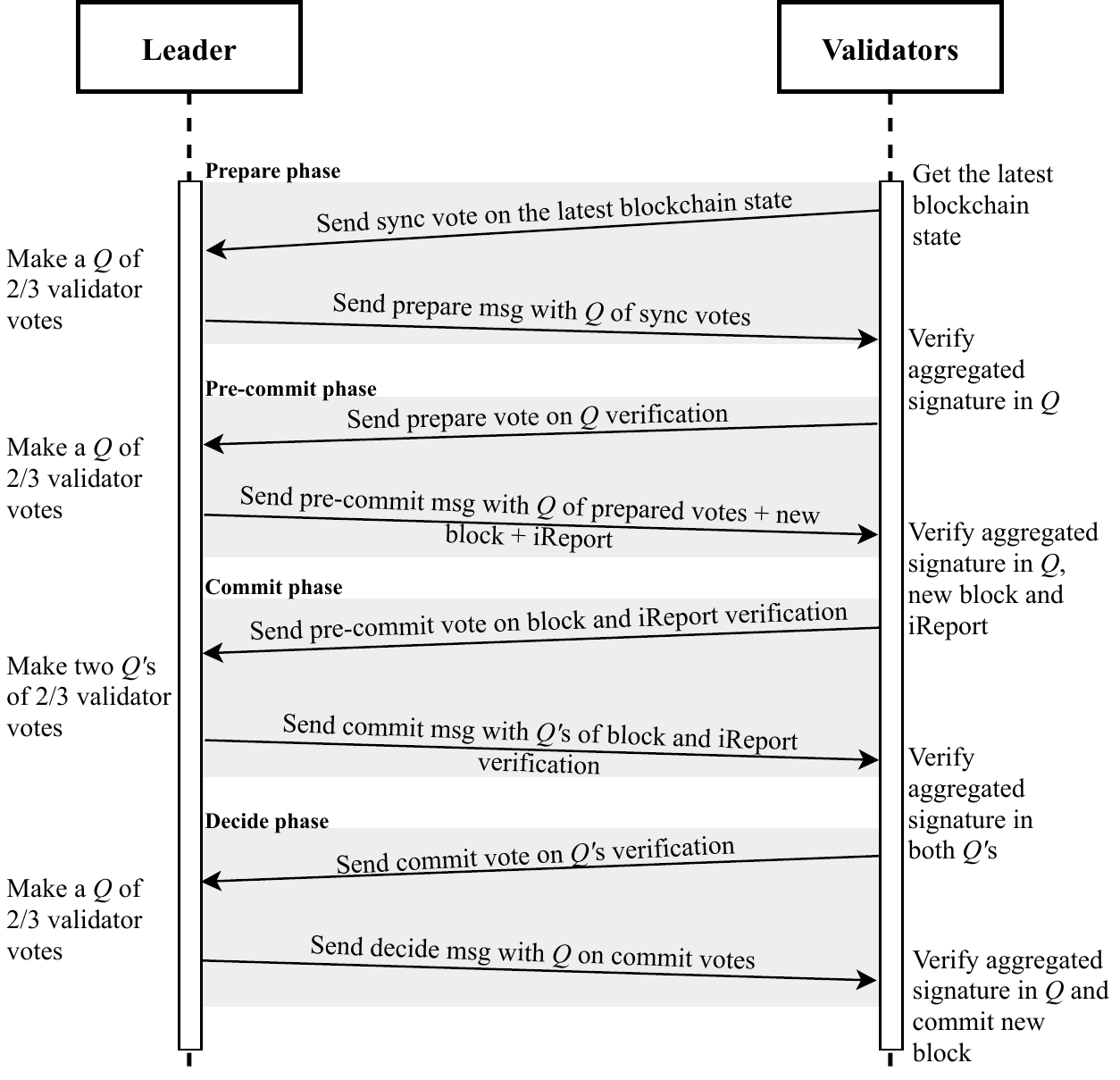}
    \caption{Communications and operations are performed between a leader and validators in a consensus run. The consensus in an NIB includes an iReport proposal and iReport verification.}
    \label{fig::consensus-algorithm}
\end{figure}

The certificate summarizes the 2/3 validator votes and includes the Ids of the voting validators, $Q_{Id}$, and a BLS aggregate signature of the voting nodes, $\Sigma_q$. {Fig.~\ref{fig::consensus-algorithm}} shows the operations and messages exchanged in a consensus run between a leader and the validators.

In the Prepare phase, the leader broadcasts $Q$ of the last block added to the blockchain. Used as a synchronization phase, validators respond with a vote after positive verification. Then, the consensus process moves to the Pre-commit phase, where the leader issues $Q$ and a block that includes the new transactions for verification. After verifying the transactions, validators respond with their votes. After a quorum is reached in the Pre-commit phase, the process moves to the Commit phase. In the Commit phase, the leader issues and broadcasts a $Q$ with the aggregated signature of the quorum on the Pre-commit phase for verification. After a quorum is reached in the Commit phase, the process moves to the Decide phase, where the leader issues and broadcast a $Q$ that includes the aggregated signature of the quorum on the Commit phase for verification. The Decide phase ends after a quorum is reached, and the block is added to the NIB blockchain.

\begin{figure}
  \centering
  \begin{subfigure}{.3\linewidth}
    \centering
    \includegraphics[scale=0.6]{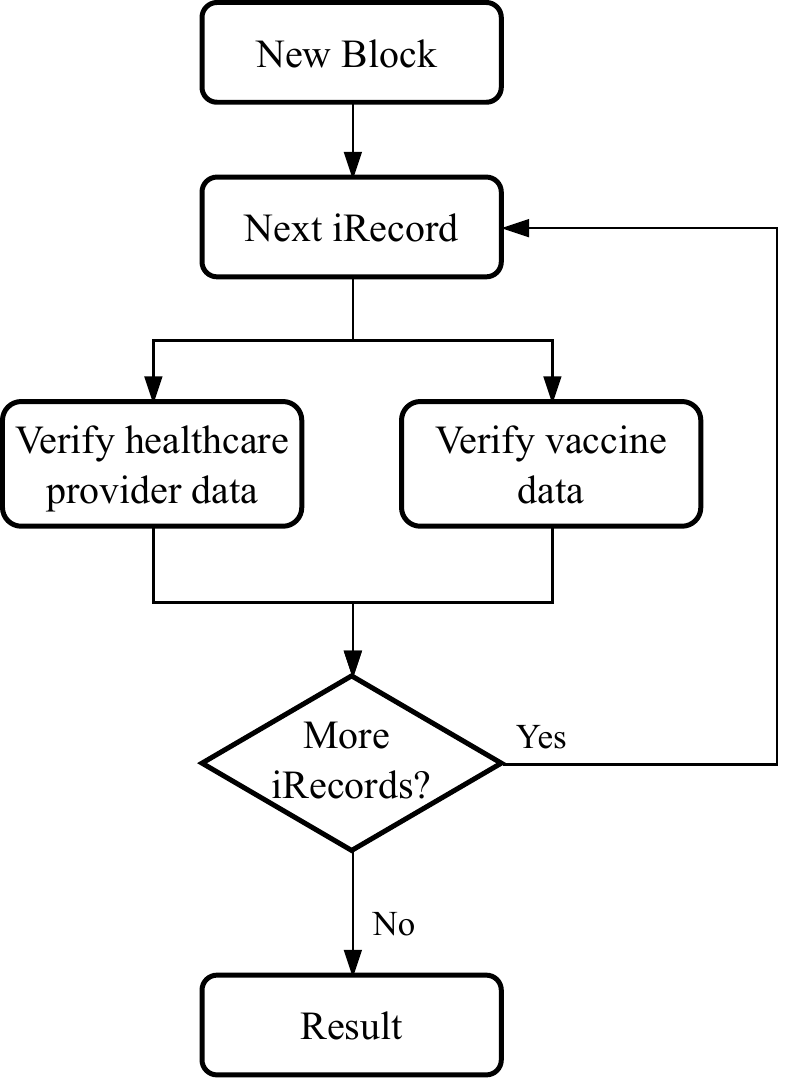}
    \caption{}
    \label{NIB_blockverification}
  \end{subfigure}%
  \hspace{8em}
  \begin{subfigure}{.3\linewidth}
    \centering
    \includegraphics[scale=0.6]{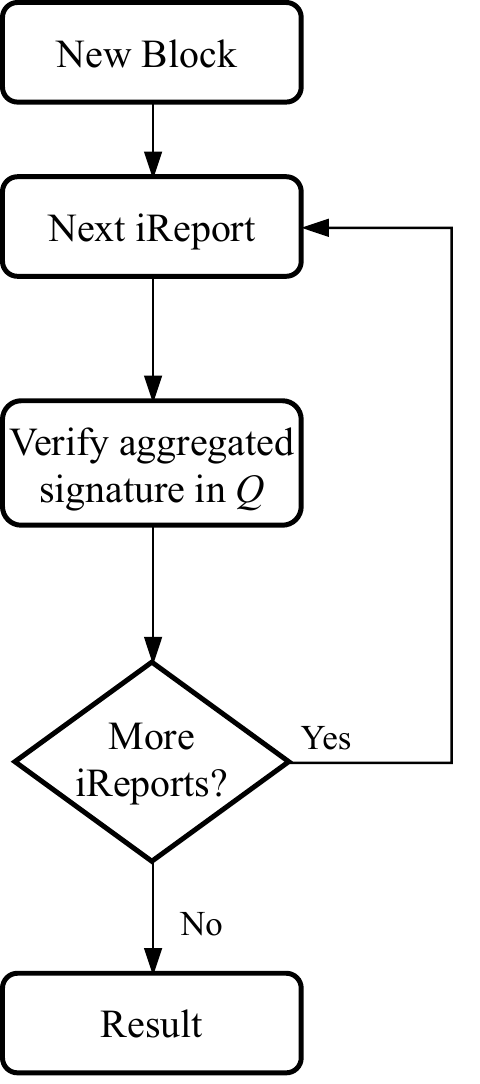}
    \caption{}
    \label{GIB_blockverification}
  \end{subfigure}%
  \caption{Block verification flow operation using (a) the vaccination application in an NIB and (b) the reporting application in GIB.}
  \label{fig::blockValidation}
\end{figure}

In an NIB, the leader also sends an iReport together with the pre-commit message to the NIB validators. The NIB validators verify the vaccination statistics of the iReport and send a vote to the leader after its verification. The NIB leader or a designated validator in the NIB sends the iReport with a $Q$ to the GIB validators for verification. GIB validators verify the aggregated signature in the $Q$ and confirm the addition of the iReport to the GIB statistical blockchain to the NIB validator that sent the initial iReport. {Fig.~\ref{fig::blockValidation}} shows the verification process of a newly proposed block in NIBs and GIB. Validators verify the transactions on that block before reaching consensus, as the figure shows.

\subsection{Recording and Accessing Vaccination Records in GEOS}
\subsubsection{Generation of iRecords and iReports}

\begin{figure}[htp!]
    \centering
    \includegraphics[width=0.6\columnwidth]{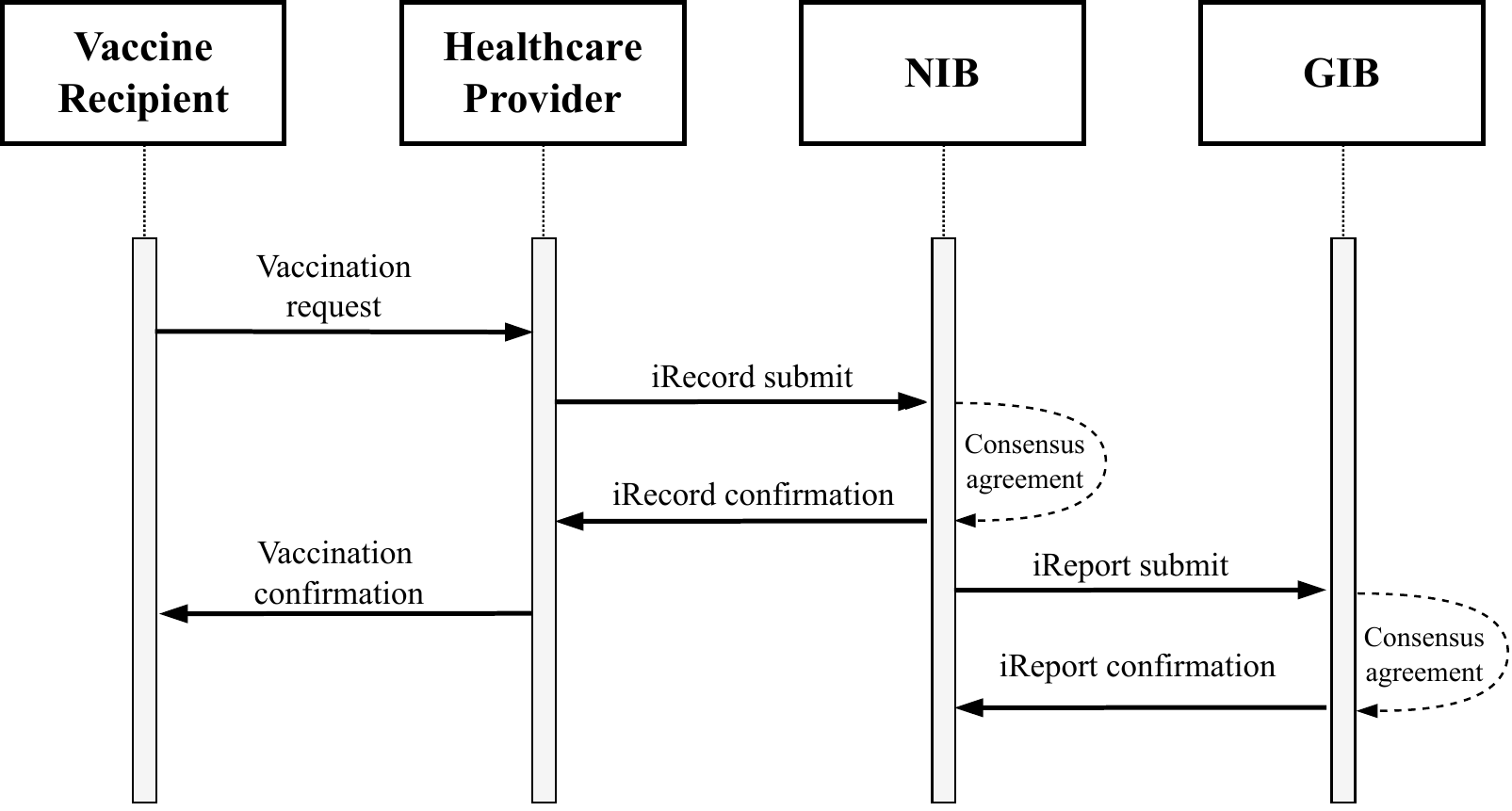}
    \caption{\small Process of adding an iRecord to NIB blockchain and issuing an iReport to GIB.}
    \label{fig::NIBRecordCreation}
\end{figure}

{Fig.~\ref{fig::NIBRecordCreation}} shows the steps for adding an iRecord to the NIB blockchain and issuing the iReport to GIB. A certified healthcare provider (i.e., a provider certified by the healthcare blockchain) confirms the identity of the vaccine recipient and administers the vaccine. After that, the healthcare provider issues an iRecord, which is sent to a validator and, in turn, to the leader. The leader includes the iRecord in a block for the following consensus process. The iRecord is committed to the vaccination repository once its block is added to the blockchain.

\subsubsection{Vaccination Verification}
\begin{figure}[htp!]
    \centering
    \includegraphics[width=0.50\columnwidth]{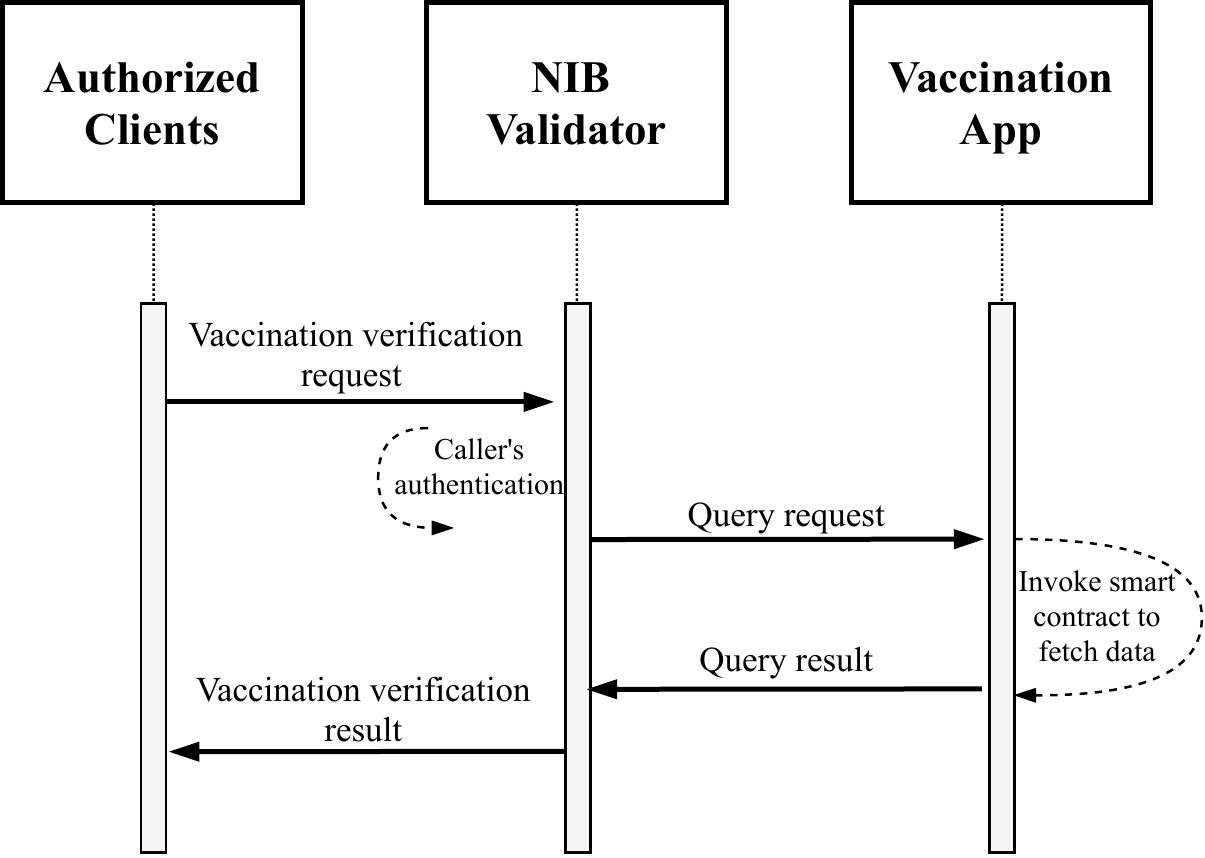}
    \caption{\small Process of vaccination verification of individuals in an NIB.}
    \label{fig::iRecordVerification}
\end{figure}

GEOS provides authorized verifiers with fast and ubiquitous access to vaccination records or iRecords. {Fig.~\ref{fig::iRecordVerification}} shows the vaccination verification process of an individual. A verification request may be for one or multiple iRecords as proof of the individual's vaccination.

\section{GEOS Confirmation Time Model}\label{sec:model}

The communications performed during the consensus process determine the confirmation times of an iRecord and an iReport and, in turn, the scalability of GEOS. In an NIB, the confirmation time of an iRecord is the elapsed period from the time when an iRecord is sent by a healthcare provider to the time when the addition of the iRecord to the blockchain is confirmed. In GIB, the confirmation time of an iReport is the elapsed period from the time an iReport is issued by an NIB validator to the time when the addition of the iReport to the GIB blockchain is confirmed. Therefore, we modeled the communications performed during the consensus process of NIB and GIB. {Table~\ref{table::Parameters}} presents the notations used to describe the confirmation time of GEOS.

\begin{table}[h]
\centering
\caption{Notations used to describe and analyze the confirmation time of GEOS.}\label{table::Parameters}
\begin{tabular}{p{25pt}p{200pt}}
\toprule
\textbf{Variable}&\textbf{Description}\\
\midrule
$s_{k_{v}}$ & Private key of a validator $v$.\\
$p_{k_{v}}$ & Private key of a validator $v$.\\
$C_g$ & A generator point in an elliptic curve.\\
$p_{k_B}$ & Group public key of validators in a blockchain.\\
$Id_{v}$ & Membership Id of a validator $v$.\\
$m_{k_v}$ & Membership public key of a validator $v$.\\
$Q$ & A quorum certificate.\\
$q_s$ & The quorum size of validator votes.\\
$\sigma_{v}$ & Signature vote of a validator $v$.\\
$\Sigma_{q}$ & A BLS aggregate signature from $q_s$ validator votes.\\
${Q}_{Id}$ & A set containing the membership Ids of signing validators in $Q$.\\
$p_{k_q}$ & An aggregated key from the public keys of a quorum of validators.\\
$B_l$& A blockchain ledger with $|B_l|$ committed blocks.\\
$b_{i}(t)$& The $i$th committed block to $B_l$ that is proposed at time $t$.\\
$b_h$& The block header, with size $|b_h|$, of a block.\\
$b_{max}$& The largest block size in bytes.\\
$R_{i}(t)$ & The set of $|R_{i}(t)|$ vaccination transactions in $b_{i}(t)$.\\
$|R|$ & The maximum number of vaccination transactions in a block.\\
$|tx|$& The size in bytes of a vaccination transaction.\\
${tx}_{ij}$ & The $j$th vaccination transaction in $R_{i}(t)$.\\
$n$& The number of validators in an NIB or GIB.\\
$\overline{|R|}$ & Average number of vaccination transactions per block.\\
${\Delta{t_i}}$ & Consensus latency of $b_{i}(t)$.\\
$\Delta{t}$ & Average consensus latency.\\
$\tau$ & Average transaction throughput.\\
$D$ & Average transaction confirmation time.\\
\bottomrule
\end{tabular}
\end{table}

The BLS scheme used in GEOS employs a pairing function or bi-linear mapping $\textbf{e}(.)$ and a hash function, $\textbf{H}(.)$, that uses the BLS12-381 pairing-friendly elliptic curve for both the mapping and the hash function {\cite{wahby2019fast}}. A validator $v$ generates a random private key, $s_{k_{v}}$, and computes its public key, $p_{k_{v}}$,  by multiplying its private key and a generator point, $C_g$, on the selected elliptic curve. Validators also aggregate their public keys to create a group public key, $p_{k_B}$, to verify the aggregated signatures in $Q$'s (\ref{eq:group-pubkey}).

\begin{align}
    &p_{k_B} = \mathlarger{\sum}\limits_{v=1}^{n}a_{v}p_{k_v}\label{eq:group-pubkey}\\
    &\qquad a_{v} = \mathlarger{\textbf{H}}\left( p_{k_v}\,, \,p_{k_1}\,||\,p_{k_2}\,\cdots||\,p_{k_{n}}\right)\nonumber
\end{align}

Before bootstrapping NIBs or GIB, validators generate their Ids and keys. Membership Ids help identify signing validators in $Q$ during the consensus process. Validators use their membership keys to vote. A validator $v$ generates its membership key, as in (\ref{eq::membershipIDs&Keys}).

\begin{align}
    &m_{k_v} = \sum\limits_{v=1}^{n}(a_{v}\cdot s_{k_{v}})\cdot \textbf{H}(p_{k_B}, Id_{v})~\label{eq::membershipIDs&Keys}
\end{align}

Let $m$ be a message holding the contents and $Q$ of a consensus phase. After verifying $Q$, a validator signs $m$ using its private key, the group public key, and its membership key, as in (\ref{eq:validatorSign}), and sends its signature (in a vote) to the leader of the consensus round.

\begin{align}
    \sigma_{v} &= s_{k_v}\cdot\textbf{H}\left(p_{k_B},m\right) + m_{k_v}~\label{eq:validatorSign}
\end{align}

A $Q$ is represented by a two-tuple $(Q_{Id},\,\Sigma_{q})$, where $Q_{Id}$ is a set holding the membership Ids of the signing validators in $Q$, and $\Sigma_q$ is the aggregated signature from the validator votes. A validator verifies $\Sigma_q$ in $Q$ using its membership Id, the group public key, and an aggregated key generated from the public keys of validators in the $Q$, $p_{k_q}$, as in (\ref{eq::QcVerification}).

\begin{align}
&\textbf{e}(C_{g}, \Sigma_{q})= \textbf{e}\left(p_{k_q}, \textbf{H}\left(p_{k_B}, m\right)\right)\cdot \textbf{e}\left(p_{k_B},\sum\limits_{v\,\epsilon\,ID_{q_v}}\textbf{H}\left(p_{k_B}, ID_{v}\right)\right)~\label{eq::QcVerification}
\end{align}

NIB and GIB blockchains use a ledger $B_l$ distributed among $n$ validators. Block $b_{i}(t)$, proposed at time $t$, is the $i$th committed block to $B_l$ (\ref{eq:blockchain-ledger}). The subscript $i$ in $b_{i}(t)$ indicates the block’s position in $B_l$. Block $b_{0}(t)$ is the genesis block that stores configuration parameters on what the validators agree beforehand. These parameters are the membership Ids and public keys of all validators, the group public key of the blockchain, the maximum block size, and the smart contracts.

\begin{align}
    B_l = \left\{b_{0}(t),\, b_{1}(t),\, \cdots,\,b_{i}(t)\right\}~\label{eq:blockchain-ledger}
\end{align}

Here, $b_{i}(t)$ contains a block header, $b_{h}$, and a set of vaccination transactions, $R_{i}(t)$. The block header contains metadata to cryptographically link the block to $B_l$ and a hash of the contents of the block to protect the transaction integrity. Validators verify every transaction in $R_{i}(t)$. Here, $tx_{ij}$ in $R_{i}(t)$ (\ref{eq::listofRecords}) denotes the $j$th transaction in $b_{i}(t)$.

\begin{align}
 & R_{i}(t) = \left\{tx_{i1},\,tx_{i2},\, \cdots\,,~tx_{ij} \right\}~\label{eq::listofRecords}
\end{align}

Validators use a memory pool, $M_{p}$, to keep incoming vaccination transactions in a First-In-First-Out (FIFO) queue. A vaccination transaction awaits in the FIFO queue until a leader in a consensus run includes it in a new block. At time $t$, the memory pool of a validator has $|M_{p}(t)|$ queued vaccination transactions. The maximum number of vaccination transactions, $|R|$, included in a new block, given the block size in bytes, is defined as a function of $b_{max}$, $|b_{h}|$, and $|tx|$, as in (7). $b_{max}$ is the largest block size indicated in the genesis block and $|tx|$, the size in bytes of a vaccination transaction. The number of vaccination transactions a leader of a consensus run includes in a block at time $t$, $|R_{i}(t)|$, is determined by the number of transactions in the validator’s memory pool at that time and $|R|$, as in (\ref{eq::numberOfTransactions}).

\begin{equation}
  |R_{i}(t)| =
    \begin{cases}
      \,\,\,|R|&\text{if $|M_{p}(t)| >|R|$}\\\\
      |M_{p}(t)| & \text{otherwise}
    \end{cases}\qquad i > 0~\label{eq::numberOfTransactions}      
\end{equation}

The average number of vaccination transactions per block, $\overline{|R|}$, in $B_l$ with $|B_l|$ committed blocks, is described as in (\ref{eq::avgTxPerBlock}).

\begin{align}
    \overline{|R|} &= \dfrac{\mathlarger{\mathlarger{\sum}}\limits_{\forall b_{i}(t)\,\epsilon\, B_l}|R_{i}(t)|}{|B_{l}|}\qquad i > 0\label{eq::avgTxPerBlock}
\end{align}

The consensus latency of $b_{i}(t)$, $\Delta t_{i}$, is defined as the time it takes for validators to reach consensus on that block. The average consensus latency, $\Delta t$, over the entire $B_l$ is defined as in (\ref{eq::avgConsensusTimes}).

\begin{align}
        {\Delta t} &= \dfrac{\sum\limits_{b_{i}(t)\,\epsilon\, B_l}\Delta t_{i}}{|B_l|}\qquad i > 0~\label{eq::avgConsensusTimes}
\end{align}

The average transaction throughput $\tau$ is the ratio of the average number of transactions per block over the average consensus latency, as in (\ref{eq::avgTransactionThroughput}).

\begin{align}
        \tau &= \dfrac{\overline{|R|}}{\Delta t}~\label{eq::avgTransactionThroughput}
\end{align}

A validator receives a vaccination transaction from a user at time $\delta t^{r}_{ij}$ and confirms receipt at time $\delta t^{c}_{ij}$, after reaching consensus on the block that includes the transaction. We calculate the average transaction confirmation time, $D$, from the confirmation times of the transactions committed to $B_l$, as in (\ref{eq::avgTxConfirmationDelay}).

\begin{align}
 &\forall\,b_{i}(t)\in B_l, \,\,\,tx_{ij}\in R_{i}(t):\nonumber\\
        &D = \dfrac{\mathlarger{\mathlarger{\sum}}\limits_{b_{i}(t)}\mathlarger{\mathlarger{\sum}}\limits_{tx_{ij}}\left(\delta t^{c}_{ij} -\delta  t^{r}_{ij}\right)}{\mathlarger{\mathlarger{\sum}}\limits_{b_{i}(t)}|R_{i}(t)|}\qquad i > 0~\label{eq::avgTxConfirmationDelay}
\end{align}

\section{Results}\label{sec:performance}

\begin{figure}[htp!]
    \centering
    \includegraphics[width=0.5\columnwidth]{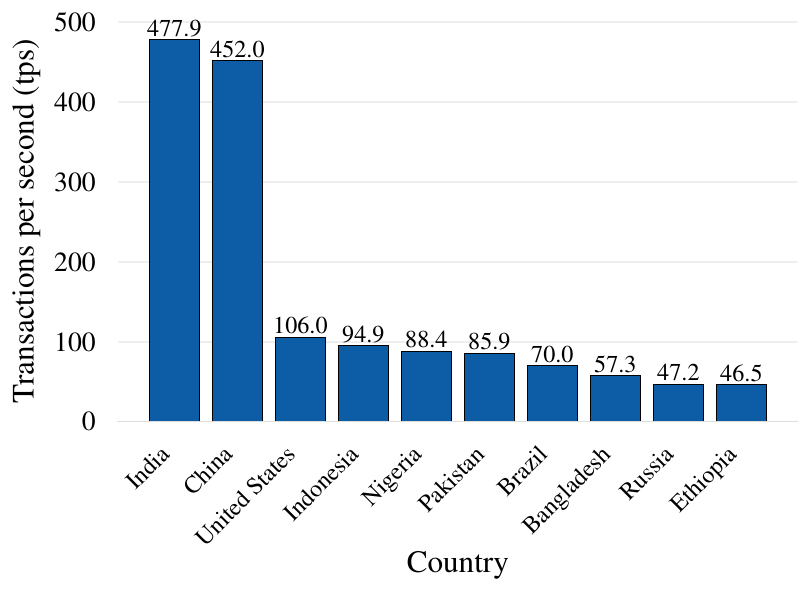}
    \caption{Top 10 countries that require the highest vaccination rates.}
    \label{fig:countryTPSEstimates}
\end{figure}

We analyze the performance and scalability of GEOS by numerically evaluating the average consensus latency, transaction throughput, and confirmation time of NIBs and GIB. We consider 184 NIBs, each corresponds to a member country of the World Health Organization (WHO). The GIB aggregates and summarizes the reports of these NIBs. Without losing generality, we considered the following assumptions in the evaluation:

\begin{itemize}
    \item The consensus algorithm uses a partially synchronous communication model and has a bounded communication delay {\cite{dwork1988consensus}}.
    \item In NIBs, the network bandwidth is 100 Mbps with a validator-to-validator propagation delay of 20 ms {\cite{speedtest2019speedtest}}.
    \item In GIB, the bandwidth is 50 Mbps and a validator-to-validator propagation delay of 120 ms {\cite{gencer2018decentralization}}.
    \item The average transaction rate for each NIB is estimated as the ratio of the number of annual administered vaccinations of the population for each country and the working hours of a day. Here, we consider 260 working days per year and 8-hour workdays. {Fig.~\ref{fig:countryTPSEstimates}} shows the top 10 countries that require the highest vaccination rate according to their population. In this context, India has the highest demand, or 478 transactions per second (tps).
\end{itemize}

\subsection{Evaluation of NIB}

\begin{figure*}[htb]
 \subfloat[]{\includegraphics[width=0.46\textwidth]{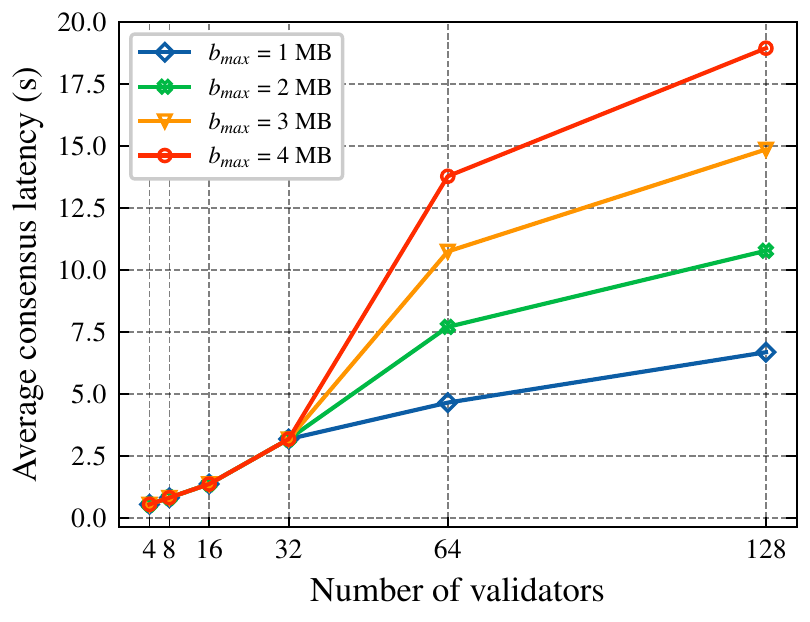}~\label{NIB_avgConsensusTime}}\hfill
 \subfloat[]{\includegraphics[width=0.46\textwidth]{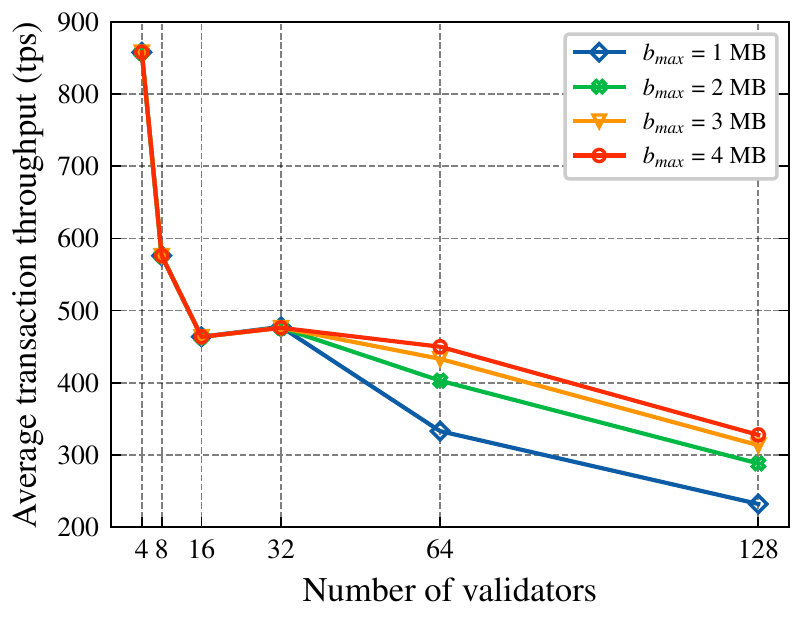}~\label{NIB_avgTransactionThroughput}}\\[-.15ex]  
 \subfloat[]{\includegraphics[width=0.46\textwidth]{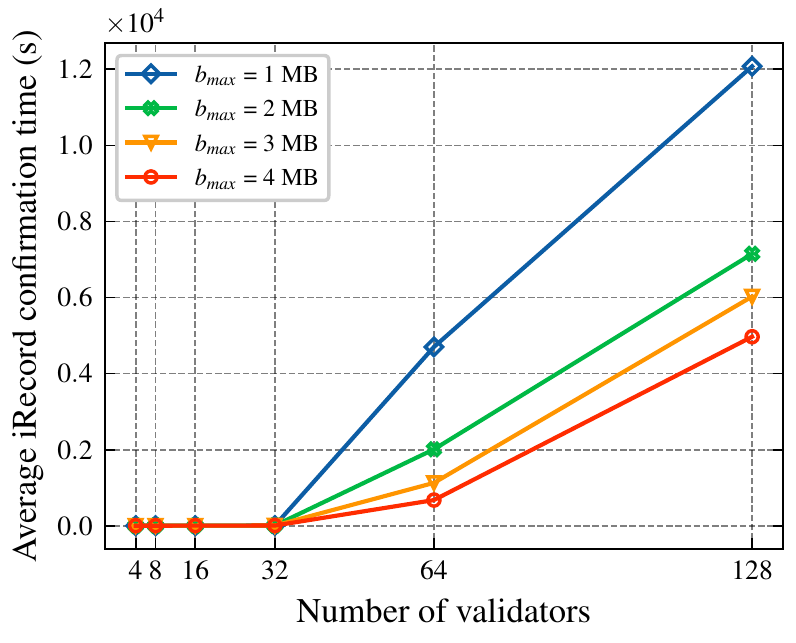}~\label{NIB_avgConfirmationDelay}}\hfill
 \subfloat[]{\includegraphics[width=0.46\textwidth]{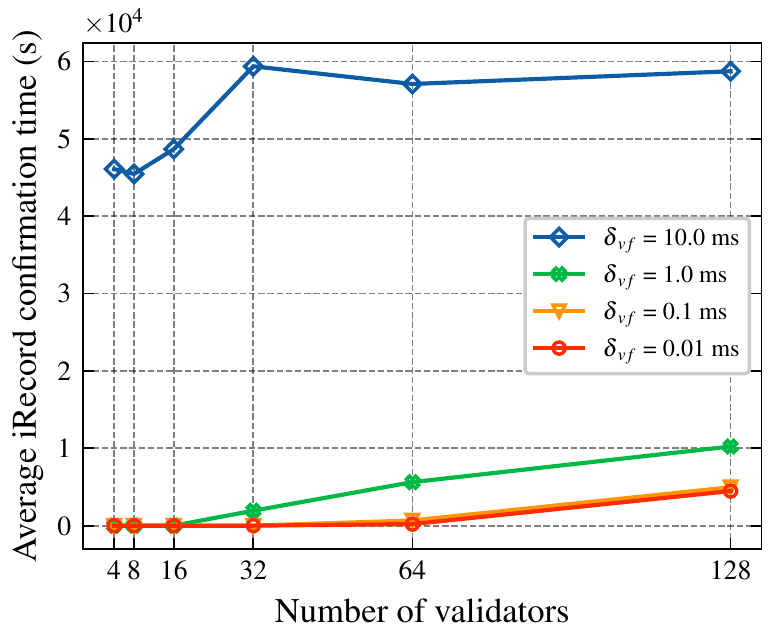}~\label{fig::NIB::verificationTimes}}
  \caption{The estimated performance of daily operations of an NIB for India with an expected vaccination transaction rate of 478 tps: (a) average consensus latency; (b) average transaction throughput; (c) average iRecord confirmation time for different block sizes; and (d) average iRecord confirmation time for different iRecord verification times, each of them as a function of the number of validators.}
  \label{NIB_results_India}
\end{figure*}

For each NIB, we assume that the generation of iRecords follows a Poisson distribution with an average arrival rate equal to the NIB’s average transaction rate and that the time to verify an iRecord is 0.1 ms \cite{gervais2015tampering}. {Fig.~\ref{NIB_results_India}} shows the evaluation results of an NIB for India.

\subsubsection{Consensus Latency}

{Fig.~\ref{NIB_avgConsensusTime}} shows that the average consensus latency increases as the number of validators and block size increase. The consensus latency of a blockchain with 128 validators is three times longer than one with 32 validators, using a block size of 2 MB. As the block size increases to 4 MB, the average consensus latency with 128 validators becomes about six times longer than one with 32 validators. We also observed that for fewer than 32 validators, the block size has no effect on the average consensus latency as the consensus latency with a few validators is small. 

\subsubsection{Transaction Throughput}

{Fig.~\ref{NIB_avgTransactionThroughput}} shows that the average transaction throughput decreases as the number of validators increases. The average transaction throughput increases as the block size increases. With the smallest number of validators (i.e., four validators), the highest achievable average transaction throughput is more than 800 tps. The minimum transaction throughput observed here is 232 tps with 128 validators and 4-MB blocks.
	
\subsubsection{iRecord Confirmation Time}
{Fig.~\ref{NIB_avgConfirmationDelay}} shows that the average iRecord confirmation time increases as the number of validators increases. With more than 32 validators, the iRecord confirmation time grows because iRecords are committed to the vaccination blockchain at a rate slower than the arrival rate of iRecords. A block size of 1 MB produces the longest average iRecord confirmation time. With such a block size, iRecords experience the longest queuing delay in the memory pools of the validators.

As the number of validators increases, the consensus algorithm slows down because the consensus requires more messages and votes. Implementing an NIB with a large block size may effectively reduce iRecord confirmation time. For example, the average iRecord confirmation time with 128 validators is longer than 3 hours, but it is reduced to about 1.38 hours when the block size is increased from 1 to 4 MB. 

We also analyze the impact of the iRecord verification time, $\delta_{vf}$, on the confirmation time. We consider different values of $\delta_{vf}$ and the number of validators and 4-MB blocks. {Fig.~\ref{fig::NIB::verificationTimes}} shows that the average iRecord confirmation time may take longer than 12 hours ($4.32\times10^4$ seconds) for $\delta_{vf}$ = 10 ms, regardless of the number of validators. However, the average iRecord confirmation time becomes less than two and three hours with $\delta_{vf}$ equal to 0.1 and 1 ms, respectively, and 128 validators. Decreasing $\delta_{vf}$ beyond 0.01 ms does not significantly reduce the iRecord confirmation time.

\subsection{Evaluation of GIB}
\begin{figure*}[htb]
 \subfloat[]{\includegraphics[width=0.46\textwidth]{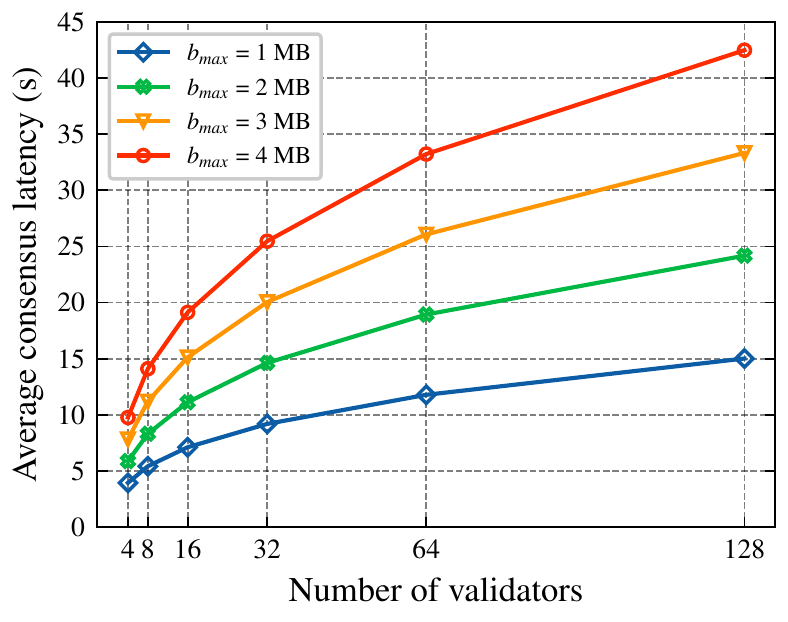}~\label{fig::GIB::ConsensusTimes}}\hfill
 \subfloat[]{\includegraphics[width=0.46\textwidth]{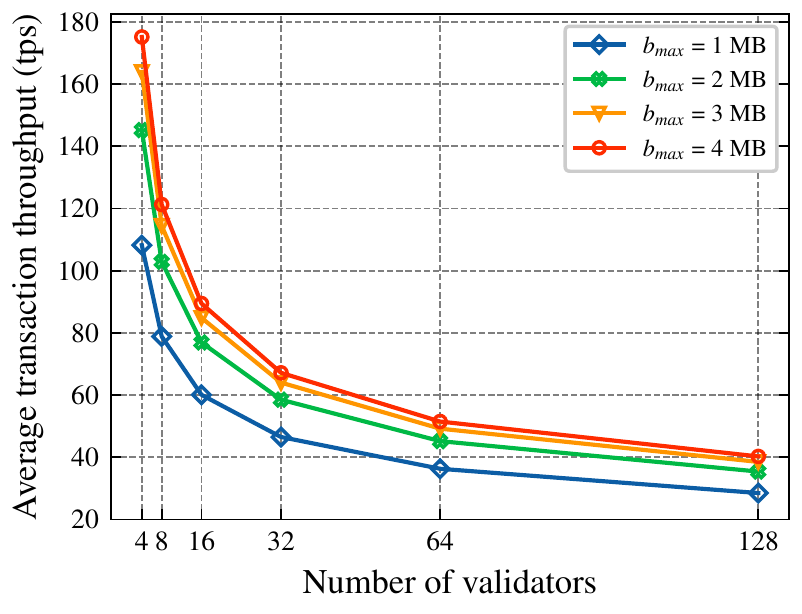}~\label{fig::GIB::TransactionThroughput}}\\[-.15ex]  
 \subfloat[]{\includegraphics[width=0.46\textwidth]{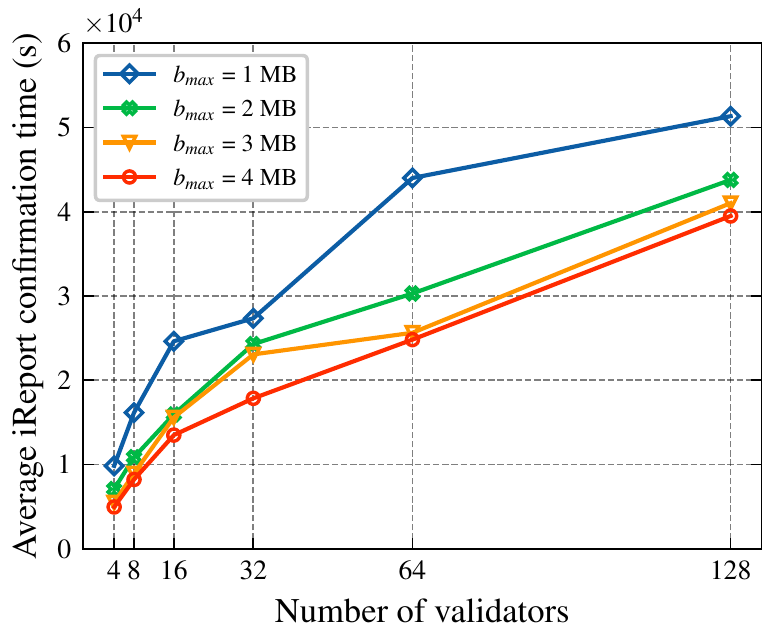}~\label{fig::GIB::TransactionConfirmationDelay}}\hfill
 \subfloat[]{\includegraphics[width=0.46\textwidth]{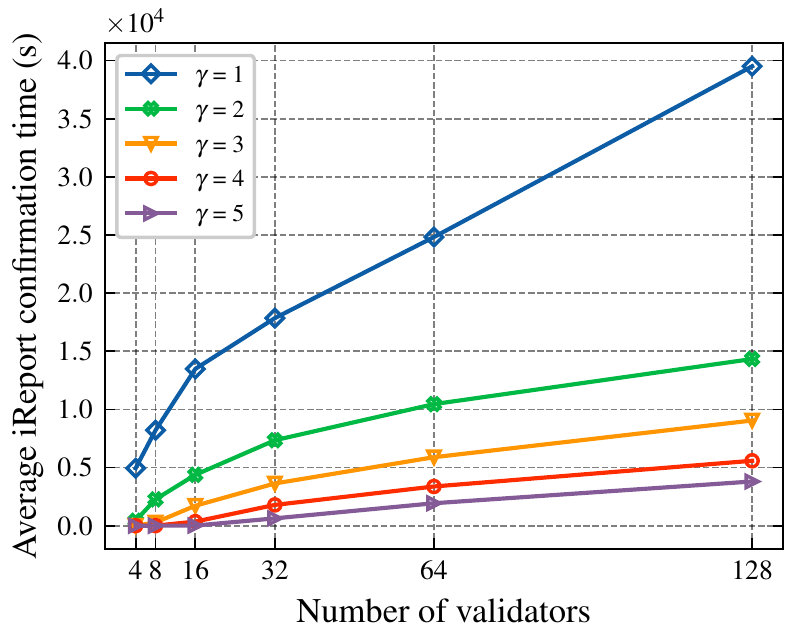}~\label{fig::GIB::sendingFrequency}}
  \caption{The estimated performance of GIB for an entire day of operations. The Figure shows the (a) average consensus latency, (b) average transaction throughput, (c) average iReport confirmation time, and (d) average iReport confirmation time with 4-MB blocks and different iReport sending rates, each of them as a function of the number of validators.}~\label{fig::TheGIBResults}
\end{figure*}

The arrival of iReports in GIB follows a Poisson distribution with an average rate equal to the sum of the average number of committed blocks per second of all NIBs. To support the vaccination rates shown in {Fig.~\ref{fig:countryTPSEstimates}}, we set the number of validators of each NIB in the range of 4 to 128. {Fig.~\ref{fig::TheGIBResults}} shows the average consensus latency, average transaction throughput, and average iReport confirmation time of GIB with 184 NIBs, as discussed below.

\subsubsection{Consensus Latency}
The average consensus latency increases as the number of validators and the block size increase, as {Fig.~\ref{fig::GIB::ConsensusTimes}} shows. Consensus in GIB takes longer than that in an NIB because the verification process of an iReport takes longer than that of an iRecord. Increasing the block size allows adding more iReports but also increases consensus latency.

\subsubsection{Transaction Throughput}
The average transaction throughput decreases as the number of validators increases, but it improves as the block size increases, as {Fig.~\ref{fig::GIB::TransactionThroughput}} shows. Increasing the number of validators increases consensus latency, decreasing transaction throughput. However, the decrease in transaction throughput may be amortized by increasing the block size. For example, when the block size increases from 1 to 4 MB, the transaction throughput increases about 38 and 29\% with 4 and 128 validators, respectively.

\subsubsection{iReport Confirmation Time}
The average iReport confirmation time increases as the number of validators increases, as {Fig.~\ref{fig::GIB::TransactionConfirmationDelay}} shows. However, it decreases as the block size increases because a larger block size allows more iReports in a block, and fewer blocks are queued. For example, the improvement achieved by increasing the block size from 1 to 4 MB is about 23\% with 128 validators. Although it is ideal for GIB to record real-time vaccination statistics, having NIBs issuing accumulated iReports as bursts may significantly increase the confirmation time of iReports. Instead, NIBs can configured to submit iReports after every $\gamma$ committed blocks, as {Fig.~\ref{fig::GIB::sendingFrequency}} shows. The figure shows that with 4-MB blocks and 128 validators, the average iReport confirmation time decreases by more than 60 and 90\% when NIBs send iReports every two and five committed blocks, respectively.

\section{Discussion}\label{sec:discussion}

In GEOS, NIBs must achieve a high transaction throughput to cope with future increases in the volume of incoming vaccinations and the number of validators. The evaluation results show that the number of validators, the block size, and the verification time dictate the latency of the consensus process and the iRecord confirmation time.
The results also show that an NIB per country may suffice to provide an iRecord confirmation time within a working day and avoid transaction queuing. Yet, the scalability of an NIB might be significantly improved by splitting it into smaller NIBs using sharding {\cite{wang2019sok,zamani2018rapidchain}}. GEOS could integrate off-chain storage solutions to reduce the amount of storage in validators {\cite{eberhardt2017or,kumar2020distributed,miyachi2021hocbs}}. Such a solution would require validators to offload committed blocks to off-chain storage while keeping hash pointers as references to such on-chain blocks. However, such an approach may also significantly increase the time to verify the pharma and healthcare data of iRecords.

\section{Related Work}\label{sec:literature-review}

Blockchain has been increasingly leading as a promising solution to different issues in e-health. Such challenges are data privacy, integrity, access, and effective and secure sharing of electronic health records; management and supervision of vaccine production, distribution, and administration~\cite{nortey2019privacy,kuo2017blockchain, clauson2018leveraging,holbl2018systematic}.

In the pharmaceutical industry, blockchain-based solutions aim to detect and prevent the counterfeiting and cloning of drugs and are proven effective in tracking the authenticity of drugs~\cite{tseng2018governance, yong2019intelligent}. Vaccine manufacturers and government health agencies may use blockchain to enhance transparency and effectiveness in vaccine production and distribution. Hu et al.~\cite{hu2019vguard} proposed vGuard, a blockchain to supervise the production and manufacturing of vaccines. Lopez et al.~\cite{ramirez2020blockchain} presented a blockchain-based management system to distribute COVID-19 vaccines.

The immutability of data records in blockchain could also ease the building of systems for monitoring and tracking emerging infectious diseases {\cite{nguyen2020blockchain}}. For example, Mashamba et al.~\cite{mashamba2020blockchain} proposed a low-cost blockchain that is an AI-coupled mobile-linked system to store individuals’ self-testing results for monitoring and tracking COVID-19 cases. Eisenstadt et al.~\cite{eisenstadt2020covid} proposed a smart-phone application that records digital certificates of COVID-19 vaccinations. In a pandemic, blockchain may also contribute as a reliable repository to consolidate information about reported cases and deaths from different reliable sources to prevent the spread of misinformation~\cite{chang2020can}. 
As a global system, blockchain has been proposed to provide real-time access, data sharing, and data privacy of medical records. The authors~\cite{biswas2021globechain} proposed GlobeChain, a blockchain architecture for real-time cross-border medical exchange services of COVID-19 data. However, the implementation and evaluation of the scalability and latency of GlobeChain are not considered, nor is the consensus algorithm for secure replication of medical records.

\section{Conclusions}
\label{sec:conclusions}

In this paper, we proposed GEOS, a scalable global immunization information blockchain-based system to reliably and securely record, track, and access information on vaccinations of the world’s population. GEOS accommodates the needs of processing the vaccination records of a country and globally.
More specifically, we modeled GEOS and the required communications in the consensus process that rules the scalability of such a global-scale blockchain. The presented numerical evaluations of the proposed blockchain model show GEOS is scalable at the country and global scales for the most demanding scenarios, such as in the case of India and the 184 countries registered in WHO. GEOS achieves an average transaction throughput larger than 300 transactions per second and an average confirmation time of up to 2 hours using both 4-MB blocks and 128 validators.

These results show that with vaccination rates similar to those used in the ongoing campaign to control the COVID-19 pandemic, GEOS can scale and provide a secure and reliable repository of vaccination records for the global population.


\bibliographystyle{unsrt}  
\bibliography{references}

\end{document}